\documentclass[aps,prl,
preprint,
showpacs,
floatfix,
]{revtex4}
\usepackage{bm}
\usepackage{amsmath}    
\usepackage{graphicx}   
\usepackage[hang]{subfigure}
\begin{document}
\title{Superconducting flux-flow type ultra-low-noise magnetic sensors\\- An alternative to dc-SQUIDs}
\author{Roberto Monaco}
\email{r.monaco@cib.na.cnr.it}
\affiliation{Istituto di Cibernetica del CNR, Comprensorio Olivetti, 80078 Pozzuoli, Italy}
\affiliation{Facolt$\grave{\rm a}$ di Scienze, Universit$\grave{\rm a}$ di Salerno, 84084 Fisciano, Italy}
\author{Carmine Granata}
\email{c.granata@cib.na.cnr.it}
\affiliation{Istituto di Cibernetica del CNR, Comprensorio Olivetti, 80078 Pozzuoli, Italy}
\author{Roberto Russo}
\email{r.russo@cib.na.cnr.it}
\affiliation{Istituto di Cibernetica del CNR, Comprensorio Olivetti, 80078 Pozzuoli, Italy}
\author{Antonio Vettoliere}
\email{a.vettoliere@cib.na.cnr.it}
\affiliation{Istituto di Cibernetica del CNR, Comprensorio Olivetti, 80078 Pozzuoli, Italy}

\date{\today}

\begin{abstract}
A superconducting magnetometers based on the magnetic field dependence of the Eck step voltage in long Josephson tunnel junctions (LJTJs) is demonstrated. The field to be measured is applied perpendicular to a continuous superconducting pickup loop. Wherever the loop has a narrow constriction, the density of the flux-restoring circulating currents will become relatively high and will locally create a magnetic field large enough to bring a biased LJTJ in the flux-flow state, i.e., at a finite voltage proportional to the field strength. This method allows the realization of a novel family of robust and general-purpose superconducting devices which, despite their simplicity, function as ultra-low-noise, wide-band and  high-dynamics magnetometers. The performances of low-T$_c$ sensor prototypes, among which a highly linear voltage responsivity and a magnetic spectral density $S_B^{1/2}< 3\,fT/Hz^{1/2}$, promise to be competitive with those of the best superconducting quantum interference devices. 
\end{abstract}

\maketitle

In the last years, high sensitive applications and sophisticated basic research experiments demanded for the development of new ultra-sensitive magnetic sensors like the atomic magnetometer based on detection of Larmor spin precession of optically pumped atoms\cite{kominis}, the hybrid magnetometer based on Giant MagnetoResistance spin valves\cite{pannetier} and the diamond magnetometer based on nitrogen-vacancy centers in room-temperature diamond\cite{taylor}. In this framework, the sensors and the circuits based on superconducting device, such as the Superconducting QUantum Interference Devices (SQUIDS), play a fundamental role, since they exhibit an extremely low noise with an equivalent energy sensitivity that approaches the quantum limit\cite{handbook,fagaly}. In this Letter, we present a novel type of superconducting magnetometers, not based on quantum interference, which combines ease of use, ultra-low-noise and high dynamic performances retaining, at the same time, the advantages of light weight, high speed and low power inherent in Josephson devices.

\begin{figure}[htb]
\centering
\includegraphics[width=8.5cm]{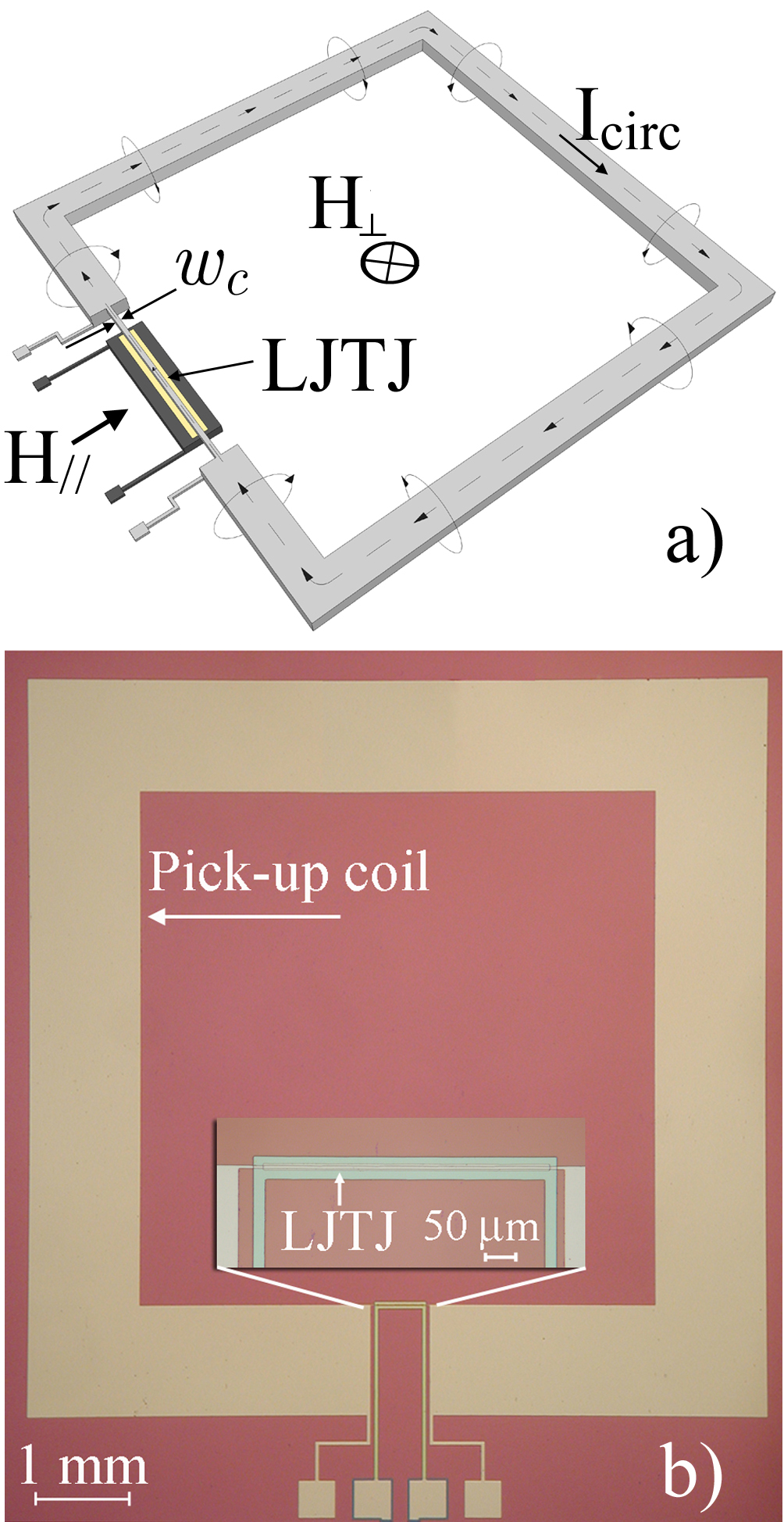}
\caption{(Color online) (a) 3D view (not to scale) of a square planar loop (light gray) with a constriction suspended over a superconducting patch (dark gray). To realize the magnetometer a window-type in-line Long Josephson Tunnel Junction (LJTJ) is realized between the loop constriction and the patch. (b) Optical image of our $Nb$-based prototype: the pickup square loop had outer dimension $D=8.0\, mm$ and width $w_p=1.1\, mm$; the loop constriction had width $w_c=5\,\mu$m and length $\ell_c=570\,\mu$m; the LJTJ, enlarged in the inset, had length ${\rm{L}}=500\,\mu$m and width $W=10\,\mu m>w_c$.}
\label{loop}
\end{figure}

Let us consider a superconducting (non-interrupted) thin-film loop with a constriction that overlaps with an insulated superconducting patch realized in a previous metalization layer. The 3D view of this system is sketched in Fig.\ref{loop}(a), where the square loop is drawn in light gray and the patch in dark gray. If a uniform (unknown) magnetic field, $H_{\bot}$, is applied perpendicular to the loop plane, then a shielding current, $I_{cir}=\mu_0 H_{\bot} A_p /L_p$, circulates in the loop to restore the initial flux\cite{mercereau}, where $\mu_0$ is the vacuum magnetic permeability, $A_p$ the effective flux capture area of the loop and $L_p$ the pickup loop inductance (we have considered that no flux was trapped in the loop during the cool down). Upon assuming the superconducting patch thicker than its magnetic penetration depth and wide enough to act as a ground plane, then an in-plane magnetic field\cite{meyers}, $H_{||} = I_{cir}/w_c =K H_{\bot}$, exists in between the constriction and the ground plane, where $w_c$ is the constriction width. By properly designing the loop and its constriction, the proportionality factor, $K=\mu_0 A_p/w_c L_p$, can be exceedingly large, that is, the constricted loop behaves as an efficient transverse to in-plane magnetic field converter. 

\noindent The unknown field, $H_{\bot}$, can be accurately determined whenever a device sensitive to $H_{||}$ is sandwiched between the constriction and the patch. Within the context of superconducting thin-films, the most obvious detector choice is a planar Josephson tunnel junction whose base and top electrodes are realized by, respectively, the patch itself and a segment of the loop. Indeed, a pioneering work\cite{PRB09} demonstrated that a jump of one magnetic flux quantum, $\Phi_0$, in the initial flux trapped in a doubly-connected electrode resulted in a small, but detectable, change of the critical current, $I_c$, of a Josephson tunnel junction. The best results were achieved\cite{SUST12} by using a one-dimensional Long Josephson Tunnel junction (LJTJ) whose width $W$ is smaller and whose length ${\rm{L}}$ is larger than the Josephson penetration length, $\lambda _J\equiv \sqrt{\Phi_0 / 2\pi \mu_0 d_m J_{c}}$, setting the junction length unit; $J_c$ is the junction critical current density and $d_m$ the junction magnetic thickness\cite{wei,SUST13}. A theoretical analysis of a DOubly-Connected-Electrode LJTJ (DOCELJTJ), corroborated by experiments, has been also reported\cite{PRB12} in which the static sine-Gordon equation\cite{ferrelOS} for an in-line LJTJ has been coupled to the quantization\cite{london} of the fluxoid in the doubly connected electrode.

\noindent In presence of a in-plane magnetic field, a LJTJ behaves like an extreme type-II superconductor\cite{barone}. The Meissner regime is reflected by a linear decrease of $I_c$ with weak magnetic fields which eventually vanishes at the critical field, $H_{c,||}=2J_c \lambda_j$, where Josephson vortices (fluxons) start to penetrate. However, accuracy of the critical current measurement depends on the switching probability (or escape rate) caused by the thermal noise. It requires the acquisition of $10^4$-$10^5$ switches by standard time-of-flight techniques\cite{fulton+castellano}, so limiting the use of the sensor proposed in Ref.\cite{SUST12} to slowly changing fields ($f<1\, Hz$). Though, for most applications a fast voltage (or current) response is mandatory. 

\begin{figure}[tb]
\centering
\includegraphics[width=10cm]{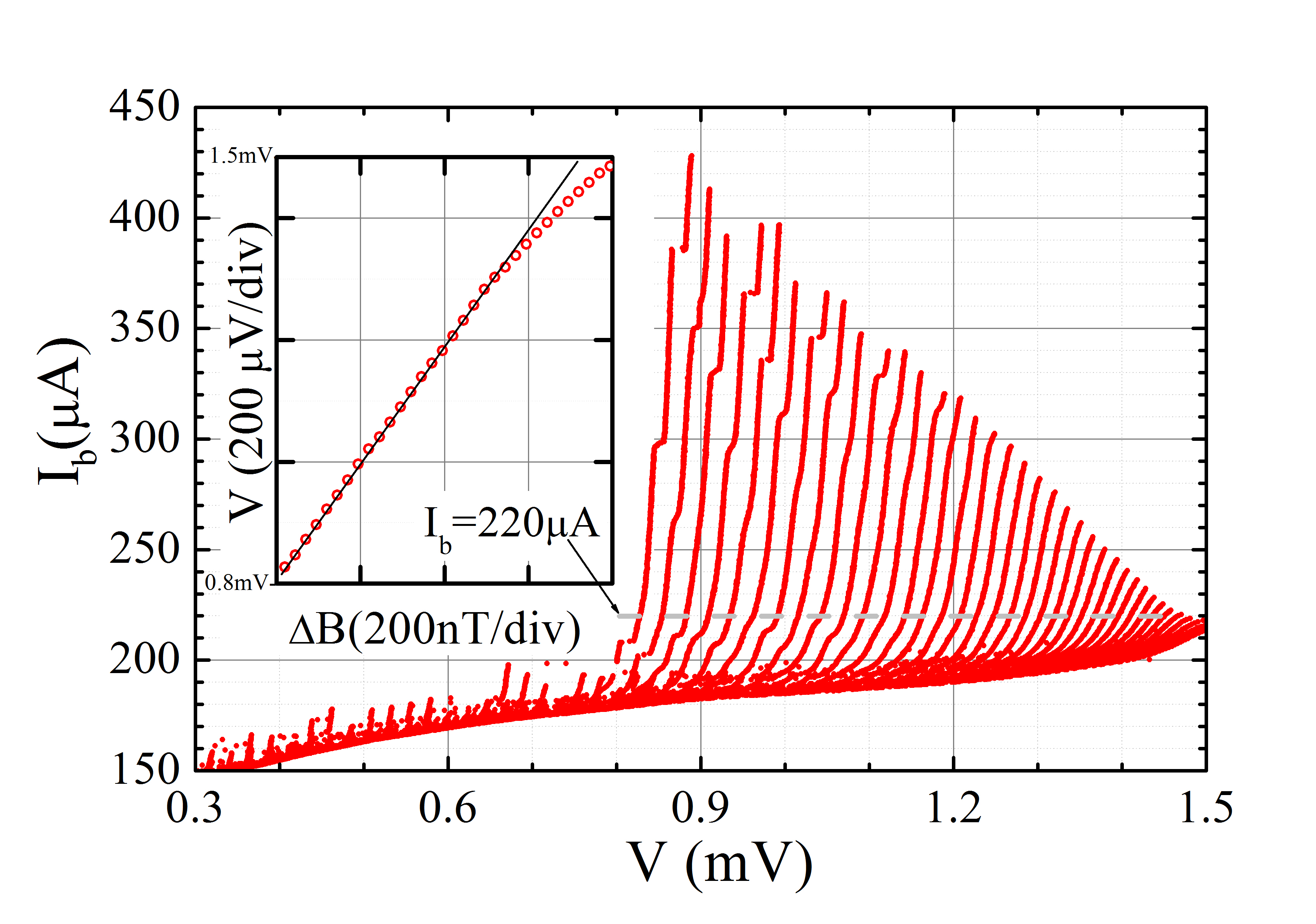}
\caption{(Color online) Current-voltage curves of a $Nb$-$Nb$ long Josephson tunnel junction measured in a purely in-line bias configuration at equally spaced values of the transverse magnetic fields. The inset shows, at a constant bias current $I_b=220\, \mu A$, the dependence of the Eck step voltage, $V_{}$, on the magnetic field changes, $\Delta B$ (open circles); the solid line helps the eye to find the linearity region. ($T=4.2\,$K)}
\label{exp}
\end{figure}

Very interestingly, a DOCELJTJ can also provide a ultra-fast and highly linear voltage responsivity offering, as will be shown, superior sensor performances. In fact, in presence of a magnetic field exceeding its critical field, a LJTJ develops a steep current singularity, called flux-flow or Eck\cite{eck} step, at finite voltages of a fraction of the gap voltage (see Fig.\ref{exp}). A flux-flow oscillator (FFO) is a LJTJ with relatively high dumping in which a unidirectional viscous flow of mutually repulsive fluxons occurs and coherent electromagnetic radiation is emitted from one of its ends where the fluxon chain collides with the boundary\cite{nagatsuma82,cirillo98}. The flux-flow in LJTJs is a very well studied phenomenon that since long is being exploited for the realization of voltage-controlled local oscillators in low noise integrated $THz$ receivers\cite{koshelets11}. The Eck step voltage, $V$, gives the number of fluxons passing per unit time across the LJTJ and is determined by two external independent stationary currents: one is referred to as the {\it control} current, $I_{ctl}$, injected into one of the junction electrodes to create the magnetic field at the two ends of the LJTJ, while the bias current, $I_b$, applied through the tunneling barrier, accelerates the fluxons and moves them from one junction extremity to the opposite one. $V$ is proportional\cite{nagatsuma82} to the in-plane magnetic field at the junction extremities, $V=\mu_0 d_m u H_{||} = \mathcal{L}_c u I_{ctl}$, where $u$ is the relativistic speed of fluxon train that cannot exceed the Swihart\cite{swihart} velocity $\bar{c}$ and $\mathcal{L}_c=\mu_0 d_m /w_c$ is the inductance per unit length of the constriction\cite{swihart,meyers}. Perfect linearity was reported in numerical simulations assuming ohmic power losses\cite{nagatsuma84,cirillo98}. The DOCELJTJ leverage is that the control current coincides with the circulating current, $I_{ctl}=I_{cir}$, which, in turn is proportional to the applied flux: in this manner, any variation of the excitation field is reflected in a linear output voltage change (at variance with a Josephson interferometer characterized by a periodic response to external flux changes). Therefore, the voltage responsivity, $V_B \equiv  \partial V /\partial B$, to a transverse magnetic field density, $B=\mu_0 H_{\bot}$, is

\begin{equation} V_B = \frac{1}{\mu_0} \frac{\partial V_{} }{\partial H_{\bot}} = \frac{1}{\mu_0} \frac{\partial V_{} }{\partial I_{cir}} \times \frac{\partial I_{cir}}{\partial H_{\bot}}= \frac{R_m A_p}{L_p},
\label{VB}
\end{equation}

\noindent where $R_m \equiv \partial V /\partial I_{ctl}= \mathcal{L}_c u$ is the so-called transresistance\cite{nagatsuma84} usual in transistor-like devices. The useful magnetic field range for the occurrence of the flux-flow state amounts to few times the transverse critical field\cite{PRB12,SUST12,SUST13}, $B_c\equiv \mu_0 H_{c,||} /K = 2 J_c \lambda_j w_c L_p/A_p \propto 1/V_B$.

The ultimate performances of any device also depend on its noise and bandwidth. To estimate the value of the minimum detectable field change, it is important to know the power spectral density, $S_B(\omega)$, of the intrinsic magnetic noise under working conditions. The noise in FFOs has been deeply studied, both analytically\cite{benabdallah,pankratov02,pankratov08} and experimentally\cite{koshelets01,valery03,KM}, since it determines the phase noise of the emitted radiation. At low frequencies the power spectral density of the intrinsic voltage fluctuation of a FFO is white\cite{mygind05}:

\begin{equation} 
S_V(\omega)\simeq S_V(0)=\left(R_d+\sigma R_m \right)^2 S_I(0),
\label{SV}
\end{equation}

\noindent where $R_d\equiv {\partial V_{}}/{\partial I_b}$ is the differential resistance of the flux-flow step, $\sigma\leq 1$ is a positive coefficient depending on the junction bias configuration\cite{koshelets01,pankratov02,pankratov12} and  
\vskip -12pt
\begin{equation} 
S_I(0)=\frac{2eI_p}{\pi}\coth\frac{eV}{k T} + \frac{eI_{qp}}{\pi}\coth\frac{eV}{2k T}
\label{SI}
\end{equation}

\noindent is the power density of the internal low-frequency current fluctuations including both thermal noise and shot noise\cite{rogovin}; $e$ is the electron charge, $k$ the Boltzmann constant, $T$ the physical temperature, $I_p$ and $I_{qp}$ are the pair and the (temperature-dependent\cite{PRB94}) quasi-particle currents, respectively. The approximation in Eq.(\ref{SV}) holds up to frequencies $\omega<< 1/R_d C$, where $C$ is the junction capacitance. We like to stress that no flicker (1/f) noise affects the flux-flow mechanism. For high-quality all-Niobium LJTJs in flux-flow state at $LHe$ temperatures, typically, the internal resistance $R_d+\sigma R_m <1 \Omega$, $I_{qp}<2I_p \approx 2 mA$ and $V_{}>2 k T/e \simeq 740\, \mu V$, therefore the voltage amplitude spectral density turns out to be $S_V^{1/2}(0)< 10 \, pV/\sqrt{Hz}$, that is by far smaller than the input noise of any available room temperature voltage amplifier ($300\, pV/Hz^{1/2}$ in a $20\, MHz$ bandwidth\cite{drung2}). Interestingly, a cryogenic electronics based on CMOS circuits showing a voltage noise as low as $50\, pV/Hz^{1/2}$ has been demonstrated\cite{much} allowing to approach the intrinsic sensor noise. Ultimately, for the proposed magnetometer the magnetic (amplitude) spectral density is

\begin{equation} 
S_B^{1/2}(0)\equiv \frac{S_V^{1/2}(0)}{V_B} =  \frac{L_p}{A_p} \left(\frac{R_d}{R_m}+\sigma \right)S_I^{1/2}(0).
\label{SB}
\end{equation}

\noindent Another source of intrinsic noise is given by the thermal noise of the loop that generates mean-square magnetic fluctuations, $B_n^2\equiv \left\langle \delta B^2 \right\rangle = k T L_p/A_p^2$, uniformly distributed in a wide frequency range ($kT/h$); it can be often neglected, especially, for larger area loops. In the experiments the low-frequency fluctuations of both the bias and control dc-currents, $I_b$ and $I_{ctl}$, unavoidably enhance the voltage noise level and should be supplied by filtered low-noise generators; however, for the stabilization of $I_{ctl}$, as it was customary in Josephson memory cells\cite{zappe80}, the persistent circulating current trapped in the loop during a proper field-cooling can be conveniently used.

The magnetometer bandwidth is upper bounded by two factors. The loop is an $R$-$L$ circuit, so the flux it encloses can change no faster than on a $L/R$ timescale; therefore, the maximum rate at which the circulating currents can follow the magnetic field variations is set by the condition $R_s(\omega_0)<\mu_0 \lambda \omega_0$, where $R_s\propto \omega^2$ is the frequency (and temperature) dependent surface resistance of the loop material (provided the loop thickness exceeds its penetration depth $\lambda$): for Niobium at $LHe$ temperature ($\lambda_{Nb} \approx 90\, nm$), we conservatively\cite{JAP00} estimated a cutoff frequency $\omega_0 \approx 2\pi \times 1\, GHz$. Another upper limit is set by the maximum rate at which the Eck step voltage can track the changes in the control/circulating current; since the intrinsic response frequency of the steady-state motion of the fluxons is typically $50$-$100\, GHz$\cite{rajeevakumar81}, this latter mechanism can be disregarded.  

As a first proof of concept demonstration, we investigated the properties of several DOCELJTJs fabricated with the tri-layer technique in which high quality $Nb/Al$-$Al_{ox}/Nb$ LJTJ were realized in a window opened in a $220$ nm thick insulating layer, so that the magnetometer pickup loop was realized at the time of the wiring layer lift-off process. Fig.~\ref{loop}(b) shows the picture of a washer-type squared loop whose outer dimension was set to $D$=8.0 mm and a width $w_p$=1.1 mm was needed to minimize the ratio $L_p/A_p$. Disregarding the constriction, the loop had a capture area $A_p=D(D-2w_p) \approx 46\, mm^2$ and an inductance $L_p=(2\mu_0/\pi)D[\ln(D/w_p)+0.5] \approx 16\,nH$, i.e., it provided a field-to-current conversion $A_p/L_p \approx 3\,\mu$A/nT. The loop constriction had width $w_c\approx 5\,\mu$m and length $\ell_c \approx 570\,\mu$m, yielding $K\simeq 720$, inductance per unit length $\mathcal{L}_c\approx 45\, pH/mm$ and total inductance $L_c= \mathcal{L}_c \ell_c \approx 26\, pH <<L_p$. The LJTJ, shown in the inset of Fig.~\ref{loop}(b), had length ${\rm{L}}\approx 500\,\mu$m and width $W\approx 10\,\mu m>w_c$, that is, it was free from  idle regions. The rectangular patch was $530\,\mu$m long, $50\,\mu$m wide and symmetrically placed with respect to the LJTJ. The external magnetic field could be applied both in the chip plane or in the orthogonal direction by means of calibrated coils. Detail of the sample fabrication and of the experimental setup can be found in  Refs.\cite{granata07,granata11}. From measurements on test junctions fabricated in the same batches, we found a Josephson current density $J_c\simeq 60\,$A/cm$^2$, resulting in a estimated Josephson penetration length $\lambda_J \simeq 50\,\mu m < {\rm{L}}/2\pi$ and a Swihart velocity $\bar{c}\simeq 1.2\times 10^7\, m/s$, corresponding to a specific capacitance per unit area of the tunnel barrier of $0.03\, F/m^2$. 

Fig.~\ref{exp} displays the current-voltage curves of the LJTJ measured with a pure in-line bias ($\sigma\approx 0.4$\cite{pankratov12}) at equally spaced values of the transverse magnetic flux density treading the pickup loop in a $800\,nT$ range.
The range lower bound depends on the magnetic flux trapped in the loop; for zero initial flux it was $1.2\,\mu T$. The Eck step approximately falls in the voltage interval $\left[V_g/3-V_g/2\right]$, where $V_g \approx 2.8$mV is the LJTJ gap voltage. The field dependence of the Eck step voltage, $V$, at a constant bias current, $I_b=220\, \mu A$ (see the gray dashed line in the main figure), is shown by the open circles in the inset where the solid line helps the eye to find the range of linearity in which we measured a voltage responsivity $V_B \simeq 900\, \mu V/\mu T$. The deviation from the linearity can be ascribed to the non-ohmic power losses\cite{cirillo98} in the Josephson barrier and/or in the junction electrodes. Therefore, the present device efficiently converts a magnetic field into a voltage with a highly linear transfer function. In the linear region the transresistance resulted to be $R_m\simeq 0.3\, \Omega$ and the step dynamic resistance varied in the range $R_d\simeq 0.3$-$0.6\, \Omega$. According to Eqs.(\ref{SV}) and (\ref{SI}), with $I_{qp}\simeq 170\, \mu A$, $I_{p}\simeq 50\, \mu A$, the expected intrinsic voltage noise is $S_V^{1/2}(0)<3 \, pV/Hz^{1/2}$. Indeed, we can only report that our room-temperature readout electronics (input voltage noise of $600\, pV/Hz^{1/2}$ in a $2\, kHz$ bandwidth), as expected, could not detect any noise originated in the device. Eq.(\ref{SB}) predicts a noise limited magnetic sensitivity $S_B^{1/2}(0)<3\, fT/Hz^{1/2}$. We stress that our design was far from being optimal and so the magnetic sensitivity is subject to further improvements, being it essentially determined by the ratios $L_p/A_p$ and $R_d/R_m$. In specie, $R_d$ can be decreased by using uniformly biased and properly tailored LJTJs\cite{benabdallah,koshelets11}, while $R_m$ increases with reduced constriction width, $w_c$, but is limited ultimately by fringing field\cite{SUST13} and quenching\cite{koch95} effects. For this sample the thermally originated, loop magnetic (integral) noise, $B_n \approx 20\, pT$, was negligible ($S_B^{1/2}=O (10^{-21} T/Hz^{1/2})$ and, throughout the operating voltage range, the dissipated power, $VI_{qp}$, was well below $1\, \mu W$.

As shown in the inset of Fig.~\ref{exp}, the linear voltage span of the Eck step exceeds $500\, \mu V$ for $Nb$-$Nb$ junctions and is independent on the loop parameters. Being the voltage range proportional to the junction gap voltage, the dynamic range would be almost doubled for $NbN$-$NbN$ samples\cite{kawakami}. For Josephson interferometers, considering the typical value of responsivity $V_{\Phi}\approx 50$-$100\, \mu V/\Phi_0$, the linear voltage span, $V_{\Phi} \Phi_0/4$, amounts to $25$-$50\, \mu V$. The wider voltage range of intrinsic linearity is the key point of DOCELJTJs, since it allows the realization of magnetometers with a highly linear and wide dynamic range (also in noisy and/or unshielded environments) and makes the use of a flux-locked loop (FFL) superfluous\cite{drung1}. In addition, the sensor bandwidth, which is often limited by the transmission line delay of the FLL circuit\cite{drung2}, can be much larger (up to several hundreds of $MHz$).

With respect to the present-day SQUID devices\cite{granata07}, the design and the fabrication process of our magnetometers are significantly simplified. In fact, they do not require shunt resistors, flux transformers, tunable resonators, modulation/feedback coils for the FLL operation and other integrated circuits, like the additional positive feedback, to increase the intrinsic responsivity. These circumstances render the fabrication yield higher and the devices more robust against the thermal cycles; in addition, the sensor operation result easier and free from parasitic effects. Indeed, the extremely low intrinsic voltage noise of the proposed sensor is a caveat which, at present, does not allow to fully benefit of its high sensitivity, as it occurred for dc-SQUIDs in the late 80's when nowadays modulated electronics was not used. This limitation can be overcome whether, for a given available detection area, the DOCELJTJ were replaced by a series array of smaller area sensors; the resulting larger internal resistance would raise the device voltage noise above a measurable threshold and, at the same time, would improve the dynamic range, the bandwidth, the field responsivity\cite{welty91,handbook} and the magnetic noise. Further, in the light of the recent progresses reported in the fabrication of multilayered high-T$_c$ planar Josephson tunnel junctions\cite{tafuri}, the DOCELJTJ low intrinsic noise is very attractive for the realization of magnetometers operating at $77\,K$ and with a very large dynamic range.

In this Letters we have discussed how a long Josephson tunnel junctions can be integrated with a superconducting loop to provide ultra-low-noise magnetometers. We stress that the proposed field detection relies on the fluxoid conservation, rather than on the Josephson interference, i.e., the DOCELJTJ design does not require any compromise between the loop and the junction parameters. In addition, its embodiment is fully compatible with most of the low- and high-T$_c$ thin film technologies developed for the fabrication of Josephson circuits\cite{hilow}. Further, the demand on the external electronics is reduced. Its performances make it particularly advantageous for the measurement of biomagnetic fields, for non-destructive testing, geological prospecting and noise thermometry. In concluding, the proposed device is seen as a natural competitor of the well-established dc-SQUID magnetometers; in perspective, it can as well be exploited for the future realization of ultra-sensitive and wide-band superconducting amplifiers.

\vskip 2pt

\noindent RM acknowledges useful discussions with V.P. Koshelets and J. Mygind.

\end{document}